\documentclass[twoside]{article}
\usepackage[latin1]{inputenc}
\usepackage{t1enc}
\usepackage{a4}
\usepackage{tabularx}
\usepackage{epsf}

\begin{document}

\begin{center}
{\Large\bf
Gamma-Ray Bursts in the 1990's -- \\[0.2cm]
a Multi-wavelengths Scientific Adventure\footnote{to appear in:
R. E. Schielicke (ed.),
Reviews in Modern Astronomy, Vol. 13 (Hamburg: Astronomische 
Gesellschaft)}}\\[0.7cm]

Sylvio Klose \\[0.17cm]
Th\"uringer Landessternwarte Tautenburg \\
D--07778 Tautenburg, Germany \\
klose@tls-tautenburg.de, http://www.tls-tautenburg.de
\end{center}

\vspace{0.5cm}

\begin{abstract}
\noindent{\it In 1997 the first optical afterglow of a cosmic Gamma-Ray 
Burst was discovered, and substantial progress
has been achieved since then. Here we present a short review of
some recent developments in this field, with 
emphasis on observational aspects of the GRB phenomenon.}
\end{abstract}

\section{Introduction}

Gamma-Ray Bursts (GRBs) are bright, transient events in the gamma-ray
sky, unpredictable in time and location, with a typical duration of
$\sim$ seconds (for a review, see Fishman 1995; 1999;  Fishman \&
Meegan 1995; Kouveliotou 1995a; Meegan 1998). The brightest bursts have 
gamma-ray fluences of order 10$^{-4}$ erg cm$^{-2}$, strong enough
to lead to detectable disturbances of the Earth's upper atmosphere
(Fishman \& Inan 1988). Most of the energy of the bursts is released
in the 0.1-1 MeV range. Spectra generally display featureless smooth
continua (for a review, see Teegarden 1998).

The first GRB was recorded with the \it Vela \rm satellites on July 2,
1967 (Klebesadel, Strong, \& Olson 1973;
Strong, Klebesadel, \& Olson 1974; for a historical review see
Bonnell \& Klebesadel 1996). The precise localization of a GRB
on the sky has been an unsolved challenge for ground-based
astronomy for 30 years, although more than 2000 bursts were
detected by numerous space-based experiments during this
timespan. This unsatisfactory observational situation led to an 
enormous flood of publications (see Hurley
1998a), and culminated in more than 100 theories about the nature of the
bursters, ranging from solar-wind
models to topological defects at cosmological distances
(in the early years of GRB research there were in fact more theories 
than bursts; see Nemiroff 1994; Ruderman 1975). The main
reason for the missing identification of the burster population was
that prior to 1997 no small, arcmin-sized GRB error boxes were available 
for rapid follow-up observations.

In the mid 1980's it was recognized that there is a subclass among the
GRBs which are distinguished from the majority of the bursts (Laros et
al. 1986, 1987). These bursts  have a soft, thermal Bremsstrahlung
spectrum ($kT$ about 30 keV), short durations, and come from certain,
well-localized regions in the sky. These \it Soft Gamma-Ray Repeaters
\rm (SGR) represent a population of objects in the Galaxy and the
Large Magellanic Cloud which occasionally emit soft GRBs. This class
of objects, currently consisting of four confirmed sources (SGR
0525--66, 1627--41, 1806--20, 1900+14), is not considered here,
although great progress was also achieved in SGR research in recent
years  (for reviews, see Hartmann 1995; Kouveliotou 1995b; Hurley
1999a).

\section{Steps toward the GRB distance scale}

\subsection{The period 1991 - 1996}

The 1990's have seen two observational breakthroughs in GRB
research. The first came with the \it Burst and Transient Source
Experiment (BATSE) \rm on the \it Compton Gamma-Ray Observatory \rm
(\it CGRO\rm; Fishman 1981; Fishman et al. 1993, 1994). It characterizes 
the period from 1991 to 1996. During this period several GRB 
experiments aboard various satellites were carried out, but \it BATSE \rm 
was most successful.

In operation since 1991, \it BATSE \rm detects about 1 burst/day,
corresponding to a full sky rate of about 800 a year (Meegan et
al. 1992).  Based on \it BATSE \rm observations it became obvious in
the early 1990's that the bursters  must be located either deep in a
Galactic halo or at cosmological distances  (Meegan et al. 1992; for
reviews, see Blaes 1994;
Dermer \& Weiler 1995; Fishman \& Meegan 1995; Harding 1994;
Hartmann
1994, 1996; Hurley 1994; Lamb 1997; M\'esz\'aros 1997). Compared to
the observational situation in the 1970's and 1980's  (for a review,
see  Higdon \& Lingenfelter 1990; Houston \& Wolfendale 1983; Puget
1981) this evidence represented substantial progress. However, it also
meant that the distance scale of the bursters was still unknown by
orders of magnitude, lying either at the kpc level or in the Gpc range
(for a detailed study see, e.g., Mao \& Paczy\'nski 1992a,b).
Correspondingly uncertain was the GRB luminosity distribution.
The arguments for and against GRBs inside or outside the Galaxy led 
to a lively debate, which was eventually presented to 
the general public (see the dedicated publications of
'The Great Debate': Fishman 1995; Lamb 1995; Nemiroff 1995;
Paczy\'nski 1995; Rees 1995; Trimble 1995).

The most promising way to distinguish between the Galactic halo model
and the cosmological model of GRBs seemed to be either to check    
with high precision the angular isotropy of the bursts on the sky
(cf. Briggs et al. 1996; Hakkila et al. 1994; Lamb 1997; Tegmark et
al. 1996b), or to search for GRBs and
their accompanying optical and X-ray flashes from a potential Galactic
halo population of bursters in M\,31 (cf. Klose 1995a; 
Lamb 1995; Li \& Liang 1992). 
Evidence for the latter was not found, but the former method
worked very well and  based on \it BATSE \rm gamma-ray data alone over
the years it became more and more difficult to postulate a Galactic
halo origin of all bursts. This included constraints on 
burst repetition (cf. Meegan et al. 1995; Tegmark et al. 1996a).

There have been several other proposals to determine the distance
scale of the bursters (cf. Paczy\'nski 1991a,b). \it At gamma-ray
energies \rm these investigations involved attempts to detect a
cosmological signature in GRB durations and spectral hardness
(cf. Brainerd 1994; Fenimore \& Bloom 1995; 
Horack, Mallozzi, \& Koshut 1996; Lee \& Petrosian 1997;
Norris et al. 1994, 1995), studies of
luminosity function effects on the observed brightness distribution of
the bursts (cf. Fenimore et al. 1993;
Horv\'ath, M\'esz\'aros, \& M\'esz\'aros 1996; Mao \&
Paczy\'nski 1992a; M\'esz\'aros \& M\'esz\'aros 1995, 1996;
M\'esz\'aros et al. 1996), analyses of the  spectral energy
distribution of the bursts in combination with pair production
opacities in GRB photon fields (cf.  Baring \& Harding 1998), and
theoretical estimates of the energy reservoir of potential Galactic
neutron stars as GRB sources (cf. Hartmann \& Narayan 1996). \it In
the X-ray band \rm a Galactic origin of the burster population would
have been revealed by scattering and absorption of GRB X-rays by the
Galactic interstellar medium (Klose 1994, 1995b; Owens, Schaefer, \&
Sembay 1995). However, this required a dedicated X-ray satellite,
which was not available at that time. \it At optical/infrared/radio
wavelengths \rm one way was to perform rapid follow-up searches for
transients following or accompanying GRBs, the occurrence of which was
predicted in theoretical models (cf. M\'esz\'aros, Rees, \& Papathanassiou
1994). Another way was to look for a potential GRB source population
in GRB error boxes. Both strategies were confronted with the problem
that \it BATSE \rm 1$\sigma$ error boxes are several degrees in radius
(cf. Briggs et al. 1999).  Although this is a very small localization
size for a gamma-ray telescope, it is a giant error box for follow-up
observations in the radio, optical, or near-infrared bands. Searches
for optical transients in GRB error boxes on wide-field
photographic plates fortuitously taken at the right time and area of
the sky have been one method to address the problem (cf. Greiner et al. 1992;
Schaefer 1990; for a review, see Hudec 1993). 
In general, this method was not very
efficient since the chance of finding suitable photographic plates
is very small (cf. Greiner et al. 1996), and variable stars could
mimic optical transients (cf. Greiner \& Motch 1995; Hudec \& Wenzel
1996; Klose 1995c; 
Pedersen 1994). Indeed, these searches did not lead to the unambiguous
detection of a GRB source. A more promising way was to set-up automatic
wide-field cameras, which could be triggered on-line by GRB detections
on board the \it CGRO\rm. Based on dedicated international
distribution networks for GRB data (cf. Barthelmy et al. 1998; Kippen
et al. 1998a; McNamara, Harrison, \& Williams 1995), response times
on the order of seconds were
achieved, however no optical transients were discovered with this
method  either (cf. Akerlof et al. 1995; Lee et al. 1997; Park et
al. 1997; Vanderspek, Krimm, \& Ricker 1995).

Another technique to observe a burster was based on GRB error
boxes provided by the Interplanetary Network (IPN) of
satellites operating in the solar system (cf. Laros et al. 1998). For
most of the 1990's the IPN consisted only of \it Ulysses-\rm
spacecraft (cf. Hurley et al. 1995) and \it CGRO\rm. The IPN makes
use of the light travel time delays between different satellites to
localize a burst on the sky (cf. Cline et al. 1980). Compared to \it
BATSE\rm-only error boxes, if triggered, the IPN can provide GRB error
boxes with considerable smaller sizes (for a review, see Lund 1995),
but not in real-time. International campaigns of follow-up
observations of the most promising half-dozen well-localized IPN
bursts only provided upper limits to the flux density of any transient
source in the X-ray, optical, or radio band (cf. Frail et al. 1994;
Galama et al. 1997a; Greiner et al. 1997a; Hurley et al. 1994; Laros
et al. 1997; Palmer et al. 1995; Pedersen 1994;
Schaefer et al. 1994; for a review,
see McNamara, Harrison, \& Williams 1995). Furthermore, deep surveys
of arcmin-sized IPN error boxes have been performed, mostly months or
years after the corresponding burst, with the aim of finding evidence
for a statistical excess of a certain class of astronomical objects in
GRB error boxes. No unambiguous GRB sources were found in the optical (cf.
Gorosabel et al. 1995; Schaefer et al. 1997; Vrba, Hartmann, \&
Jennings 1995; Webber et al. 1995), although in the case of quasars
and active galactic nuclei the situation is still somewhat
controversial (cf. Burenin et al. 1998; Fruchter et al. 1999b; Hurley
et al. 1999a; Luginbuhl et al. 1995, 1996; Schartel, Andernach, \&
Greiner 1997; Vrba et al. 1994, 1999a).  
Similar surveys in the radio
(cf. Palmer et al. 1995; Schaefer et al. 1989), the X-ray band
(cf. Bo\"er et al. 1993, 1997; Greiner et al. 1991;
Hurley et al. 1999b; Pizzichini et
al. 1986), and the near-infrared (Blaes et al. 1997;  Klose,
Eisl\"offel, \& Richter 1996; Klose et al. 1998; Larson \& McLean
1997; Larson, McLean, \& Becklin 1996) also failed to identify the
burster population. The deepest optical survey (Vrba, Hartmann, \&
Jennings 1995) and the near-infrared surveys revealed, however, that
there is no lack of potential GRB host galaxies in IPN error boxes.
The missing item was the direct observational proof of an
extragalactic origin of the bursts. 
This second observational breakthrough was initiated by the
Italian-Dutch gamma-ray/X-ray satellite \it BeppoSAX \rm launched in
1996 (Boella et al. 1997; Costa et al. 1998; Frontera 1998),
and later it included successful GRB localizations with RXTE 
(see Bradt \& Smith 1999; Smith et al. 1999). It
characterizes the period from 1997 to 1999. The observational
situation before that time is summarized by e.g.  Greiner (1995), Lamb
(1997), McNamara, Harrison, \& Williams (1995), M\'esz\'aros (1997),
and Vrba (1996).

\subsection{The period 1997 - 1999}

Contrary to \it BATSE, \rm \it BeppoSAX \rm can provide an
arcmin-sized GRB error box within hours after the GRB trigger,
though only about ten times a year because of its small field of
view. The first burst detected by \it BeppoSAX \rm in 1997 was GRB
970111 with an error box of 10 arcmin radius provided to
observatories via electronic mail only hours after the occurrence of
the burst (Costa et al. 1997a). Although no GRB afterglow was detected
either in the optical (cf. Galama et al. 1997b) or in the radio band
(Frail et al. 1997a), this burst made it immediately clear that an
observational breakthrough was near. It came with GRB 970228 (Costa et
al. 1997b,c) when for the first time the optical afterglow of a burst
was discovered (Groot et al. 1997; van Paradijs et al. 1997), meaning
that the burst was localized with high angular precision. Since then 
much has been learned about GRBs based on multi-wavelengths observations
of afterglows. The establishment of a sophisticated electronic
network for GRB messages, the \underline{G}RB \underline{c}oordinated  
\underline{n}etwork GCN (Barthelmy et al. 1998, see appendix), was another
milestone on this route. Various aspects of these exciting, new
discoveries have been summarized by numerous authors from the
theoretical as well as from the observational point of view  (e.g.,
Blinnikov 1999; Castro-Tirado 1999; Frontera 1998; Ghisellini 1999;
Greiner 1998; Hartmann 1999; Hurley 1998b; Lamb 1999; McNamara \& Harrison 
1998; M\'esz\'aros 1999a,c; M\'esz\'aros, Rees, \& Wijers 1999; Paczy\'nski
1999; Piran 1999a,b; Rees 1998, 1999; Stern 1999; Vietri 1999).

\section{Recent observations of GRB afterglows \label{afterglow}}

The occurrence of broad-band afterglows following GRBs was expected
on theoretical grounds (cf. Rees 1998, and references
therein). Compared to the duration of the bursts, afterglows in the
long-wavelengths bands can be long-lived, making the precise
localization of the bursters possible and extending GRB research into
international multi-wavelength observing campaigns. Greiner's
WWW-page at the Astrophysical Institute Potsdam (see appendix)
provides informations about ongoing observational activities in GRB
follow-up studies.

According to the currently most accepted theoretical GRB model, the 
afterglows are due to external shocks when a relativistically
expanding fireball (a $\gamma$-ray fireball, see Cavallo \& Rees 1978)
released by a compact source sweeps up matter from
the ``interstellar'' medium surrounding the burster (for an introduction
into this subject and/or a review, see M\'esz\'aros 1997, 1999a; Piran
1997, 1999a,b; Rees 1999). This medium could be, for example, the
ordinary interstellar medium in a spiral galaxy, or the stellar wind
environment from the GRB progenitor (cf. Chevalier \& Li 1999; Halpern
et al. 1999). The afterglow emission process is most likely synchrotron
radiation (see, e.g., Sari, Piran, \& Narayan 1998; Wijers, Rees, \&
M\'esz\'aros 1997, and references therein).

At the date of submission of this review, long-lasting 
optical afterglows have been observed from about a dozen
GRBs, and about ten radio afterglows were discovered (see 
Greiner's WWW-page [appendix]; 
see also Table 2 in Wheeler 1999). Evidence for a short-lived
afterglow in the gamma-ray band has been reported too (Giblin et
al. 1999). About half of all GRB afterglows seen in the X-ray band
showed no detectable optical emission, whereas almost all
GRBs detected by \it BeppoSAX \rm exhibited an X-ray
afterglow (Costa 1999).

To date, there are eleven reported spectroscopic redshift measurements of 
GRB afterglows and/or host galaxies (mid December 
1999; Table 1), confirming the cosmological distance scale of GRBs, which 
was considered soon after their discovery (e.g., Usov \& Chibisov 1975;
van den Bergh 1983; Paczy\'nski 1987). The data seem to indicate that the
redshift distribution of the bursters peaks around 1, with a long
tail towards higher redshifts (for model fits see, e.g., Schmidt
1999). There is also one burst that came from the local universe
($z$=0.0085; GRB 980425). The question if all GRBs are of
extragalactic nature is not yet completely solved, however (cf. Cline,
Matthey, \& Otwinowski 1999; Tavani 1998).

\begin{table*}[t]
\caption{\bf Gamma-ray bursts with reported spectroscopic redshifts\rm}
\vspace*{3mm}
\begin{tabular}{c|l|r|r|r|c|c}
\hline
GRB                 &
$z$ $^{\ a}$        &
Ref.                &
$F_\gamma^{\ b}$    &
band$^{\ c}$        &
Ref.                &
$E_\gamma^{\ d}$    \\
\hline
\hline \\[-2mm]
970228 & 0.695 &  1 & 2          & 35...1000& 11 & $2.4\,\times\,10^{51}$ \\
970508 & 0.835 &  2 & 3          & 20...1000& 12 & $5.3\,\times\,10^{51}$ \\
970828 & 0.96  &  3 &            &          &    &                        \\
971214 & 3.418 &  4 & 11 $\pm$ 1 & $>20$    & 14 & $3.5\,\times\,10^{53}$ \\
980425 & 0.0085&  5 & 4 $\pm$ 1  & $>20$    & 15 & $7.4\,\times\,10^{47}$ \\
980613 & 1.096 &  6 & 2          & $>20$    & 16 & $6.1\,\times\,10^{51}$ \\
980703 & 0.966 &  7 & 30 $\pm$ 10& 40...700 & 17 & $7.1\,\times\,10^{52}$ \\
990123 & 1.600 &  8 & 509 $\pm$ 2& $>20$    & 18 & $3.4\,\times\,10^{54}$ \\
990510 & 1.619 &  9 & 26 $\pm$ 1 & $>20$    & 19 & $1.8\,\times\,10^{53}$ \\
990712 & 0.430 & 10 &            &          &    &                        \\
991208 & 0.707 & 20 &            &          &    &                      \\[0mm]
\hline
\end{tabular}
\vspace*{2mm}
\newline  $^a$ redshift;\ $^b$ observed gamma-ray fluence  [$10^{-6}$
          erg cm$^{-2}$] based on measured photon flux; 
          \ $^c$ corresponding energy band [keV]; \
          $^d$ released gamma-ray energy [erg], isotropic  emission
          assumed and calculated for a standard Friedmann cosmology
          with $H_0=65$ km s$^{-1}$ Mpc$^{-1}$, $\Omega_0=0.2$
\newline \small
\newline References:
 1:  Djorgovski et al. 1999, GCN 289; \,
 2:  Bloom et al. 1998, ApJ 507, L25; \,
 3:  Frail 1999, 5th Huntsville meeting on GRBs; \,
 4:  Kulkarni et al. 1998, Nature 393, 35; \,
 5:  Tinney et al. 1998, IAU Circ. 6896; \,
 6:  Djorgovski et al. 1999, GCN 189;\,
 7:  Djorgovski et al. 1998, ApJ 508, L17; \,
 8:  Keson et al. 1999, IAU Circ. 7096; Hjorth et al. 1999, GCN 219; \,
 9:  Vreeswijk et al. 1999, GCN 324; \,
10:  Galama et al. 1999, GCN 388; \,
11:  Palmer et al. 1997, IAU Circ. 6577; \,
12:  Kouveliotou et al. 1997, IAU Circ. 6660; \,
14:  Kippen et al. 1997, IAU Circ. 6789; \,
15:  Kippen et al. 1997, GCN 67; \,
16:  Woods et al. 1998, GCN 112; \,
17:  Amati et al. 1998, GCN 146; \,
18:  Kippen et al. 1999, GCN 224; \,
19:  Kippen et al. 1999, GCN 322; \,
20:  Dodonov et al. 1999, GCN 475. \normalsize
\end{table*}

Usually the optical transient following a GRB has an $R$-band
magnitude of about 18...22 when it is detected some hours after the
burst, provided that no strong extinction occurs in the GRB host
galaxy or in our Galaxy. This can make the optical transient
detectable to 1-m class telescopes. 
For example, the optical afterglow of GRB 970508 was
discovered with the Kitt Peak 0.9-m reflector (Bond 1997), the
afterglow of GRB 980329 with the Tautenburg 1.34-m (Klose, Meusinger,
\& Lehmann 1998; Reichart et al. 1999), and the afterglow of GRB
990308 with the Venezuelan 1-m Schmidt telescope (Schaefer et
al. 1999). The burst 980329 is the only GRB afterglow detected  in
the submm-band (Smith et al. 1999). Presumably, there was something
special about this burst, which is also indicated by the color of its
afterglow (for a discussion see, e.g., Draine 1999; Fruchter 1999; 
in 't Zand et al. 1998; Palazzi et al. 1998). Detailed color
measurements of afterglows are published for, e.g., GRB 980703
(Vreeswijk et al. 1999) and 990510 (Israel et al. 1999).

The time-dependent flux density of an afterglow follows a power-law
decay, $F_\nu (t) \sim t^{-\beta}$, in accordance with fireball models
(cf. Sari, Piran, \& Narayan 1998; Wijers, Rees, \& M\'esz\'aros
1997). For example, a decay constant in the optical
of $\beta=1.2$ leads to a
dimming of the optical transient by 3 photometric magnitudes  between
day 1 and day 10 after the occurrence of the burst. Rapid follow-up
observations are therefore crucial for the detection of GRB
afterglows. Exceptions to this rule do occur, however. The afterglow
flux of GRB 970508 showed a fall and rise during the first 2 days,
before it entered a power-law decline (cf. Castro-Tirado et al. 1997;
Fruchter et al. 1999b; Pedersen et al. 1998; for a possible
theoretical explanation, see Panaitescu, M\'esz\'aros, \& Rees 1998;
Pugliese, Falcke, \& Biermann 1999).

In its early phase a GRB afterglow can outshine its host galaxy, which
becomes visible weeks or months after the burst when the light curve
of the afterglow seems to flatten. This was first seen for GRB 970508
(Zharikov \& Sokolov 1999). This was also the first burst where a
radio counterpart of the afterglow was detected with the 
Very Large Array (VLA) which
suggested extreme relativistic expansion velocities of the radio
source (Frail et al. 1997b; Waxman, Kulkarni, \& Frail 1998) based on
the radio scintillation technique (Goodman 1997). Meanwhile, it has
been proven that observations in the radio band can localize GRB
afterglows in \it BeppoSAX \rm error boxes when optical
identifications are still missing (Taylor et al. 1998a,b, in case of
GRB 980329; Frail et al. 1999a, in case of GRB 991208) 
and even when they do not exist at all (Frail 1999, in
case of GRB 970828; Frail et al. 1999b, in case of GRB 981226).

\section{Afterglows and GRB energetics \label{beaming} }

Perhaps the most impressive property of the bursts is the huge amount
of electromagnetic energy released (Table 1). This is especially 
apparent in the case of GRB 990123, where an optical flash was
detected for the first time when the burst was still in progress in the
gamma-ray band (Akerlof et al. 1999). This phenomenon was predicted
from theory (M\'esz\'aros 1999a, and references therein; Sari \&
Piran 1999, and references therein; for a phenomenological approach,
see Ford \& Band 1996). The optical flash peaked about
45 sec after the onset of the burst at a mean $V$-band magnitude of
about 9 on a frame with a 5 sec exposure (Akerlof et al. 1999). The
(minimum) redshift of the burster was found to be 1.60 (Andersen et
al. 1999; Kulkarni et al. 1999), so that the optical flash translates
into an ultraviolet rest-frame luminosity of about $3\,\times\,
10^{16} \ L_\odot$ (Kulkarni et al. 1999). If the burster had been at a
distance corresponding to that of M\,81, its $V$-band magnitude
would have reached \mbox{--9} (a cosmological $K$-correction
neglected). If it had been located at the Galactic center, but assuming 
no extinction, it would have peaked at a $V$-band magnitude of
\mbox{--22}, comparable to the apparent magnitude of the Sun. Assuming
isotropic emission, the gamma-ray energy emitted by this burst amounts
to $3.4\,\times\,10^{54}$ erg (Table 1), 
about twice the rest mass energy of the
Sun and an order of magnitude higher than the previous record holder,
GRB 971214 (Kulkarni et al. 1998a). The term 'hypernova' (Paczy\'nski
1998a,b) has become a popular word to describe such energetic events.

Two basic ideas are being discussed on how to reduce the
required energy budget; gravitational lensing and non-isotropic
emission. This discussion peaked when GRB 990123 was detected and its
redshift measured (see the archive of the GCN circulars; appendix).
Direct observational support for lensing has never been found in GRB data
(cf. Marani et al. 1999) but evidence for beaming seems to be present.
Beaming reduces the energy requirement by a factor of order 10--100,
depending on the assumed or calculated solid angle of emission, but it
increases by the same factor the required GRB event rate.

It was recently predicted that beaming of the relativistic
outflow  should lead to a break in the afterglow light curve
(M\'esz\'aros \& Rees 1999b; Panaites\-cu, M\'esz\'aros, \& Rees 1998;
Piran 1999b; Rhoads 1997, 1999; Sari, Piran, \& Halpern 1999) and to a
non-zero and time-dependent polarization in the optical/near-infrared
(Ghisellini \& Lazzati 1999; Gruzinov 1999; Sari 1999). This is
expected to occur when the decreasing bulk Lorentz-factor of the
radiating shock front which runs into the  ambient interstellar medium
of the burster allows the observer to see the edge of the jet (Piran
[1999b] notes that 'jet' is not the appropriate description of the
phenomenon, since this is a transient collimated outflow). The
smaller the opening angle of the jet, the earlier the observer
should see a break, a steepening in the afterglow light curve. Within
this theoretical context it would be interesting to learn if a
time-delayed signal from the counterjet is expected to be seen too.

Observational evidence for beaming was first found in the optical
afterglow of GRB 990123, where the light curve steepened two days
after the burst (Castro-Tirado et al. 1999; Kulkarni et al. 1999). The
measured steepening is in agreement with the picture that at this time
the observer began to see the edge of a jet. Recently, a steepening in
the light curve has also been observed for another burst, GRB 990510
(Beuermann et al. 1999; Harrison et al. 1999; Stanek et al. 1999). In
both cases a beaming factor in the order of 200...300 has been
deduced  from the observations (Harrison et al. 1999; M\'esz\'aros
1999b; Sari, Piran, \& Halpern 1999). This seems to bring very
energetic bursts to the energy output of less energetic ones (Table
1), for which no  observational evidence for strong beaming has been
found, like GRB 970228 and 970508 (Sari, Piran, \& Halpern 1999). 
More afterglow observations are required, however, to check this
hypothesis.

An exciting consequence of beaming is that there could exist GRBs
which develop an X-ray, optical, or radio afterglow, but have no
detected gamma-ray burst (for a discussion see, e.g.,  M\'esz\'aros,
Rees, \& Wijers 1998, 1999; Perna \& Loeb 1998b; Rhoads 1997). Archived X-ray
data have been  searched for such events, but no strong evidence for
them was found (Greiner et al. 1999a,b; Grindlay 1999).
However, possible evidence for a relation of optical transients detected
on photographic plates to underlying blue galaxies 
was reported by Hudec et al. (1996). Results of wide-field CCD surveys 
to search for this phenomenon have not yet been published.

There are numerous possible causes for a break in the light curve of
an afterglow (for a discussion see, e.g., Kulkarni et al. 1999; Wei \&
Lu 1999). An independent observational test for beaming is therefore
desirable. It could be provided
by polarimetric observations. According to recent theoretical studies,
optical afterglows could be linearly polarized up to the 10\% level if
the radiation comes from a collimated outflow moving with relativistic
velocity toward the observer and the observer is not directed exactly
at the center of the jet (Ghisellini \& Lazzati 1999; Gruzinov 1999;
Sari 1999; see also Hughes, Aller, \& Aller 1985).  However,
microlensing is also considered as a possible option for producing
time-dependent linearly polarized afterglows (Loeb \& Perna 1998a). If
the degree of linear polarization of an afterglow can be as high as
predicted, even if no beaming occurs (Gruzinov \& Waxman 1999), this
could allow observers to detect an afterglow in a GRB error box  at
high Galactic latitude without the need of second-epoch data. This
holds particularly in the near-infrared bands (Klose, Stecklum, \&
Fischer 1999).

Linear polarization was first detected at the 2\% level in the
afterglow of GRB 990510 about 1...2 days after the burst based on
observations with the ESO Very Large Telescope (Covino et al. 1999; Wijers et
al. 1999). Unfortunately, only three data points could be obtained, and within 
the measurement errors no time-dependency of the degree of linear
polarization was found. However, such a time-dependence appears to be 
present in the afterglow of GRB 990712 (Rol et al. 1999).

\section{What is the nature of the bursters? \label{extinction} }

Among the most exciting discoveries about the nature of the GRBs is
the bimodality of their duration distribution (Kouveliotou et
al. 1993; Mazets et al. 1981; McBreen et al. 1994). Recently,
evidence for an intermediate class of bursts was  also reported
(Horv\'ath 1998; Mukherjee et al. 1998). Short burst durations range
between 0.01 to 2 seconds, long bursts last from about 2 to a few
hundred seconds. Since \it BeppoSAX \rm is only sensitive to long
bursts, it is currently unknown whether short bursts do also produce
detectable GRB afterglows. It is also unknown what the origin of this
bimodality is although various studies have been performed to gain 
insights into this issue (cf. Bal\'azs, M\'esz\'aros, \& Horv\'ath 1998;
Bal\'azs et al. 1999; Dezalay et al. 1996;
Katz \& Canel 1996; Mao, Narayan, \& Piran 1994;
M\'esz\'aros, Bagoly, \& Vavrek 1999; Mitrofanov 1998).

One working hypothesis is that the bimodal duration
distribution reflects different GRB engines. According
to the most accepted picture of the bursters today, GRBs can be made
either by merger events that include compact stars and/or stellar-mass
black holes, or by the gravitational collapse of single stars
(cf. Fryer \& Woosley 1998; Fryer, Woosley, \&
Hartmann 1999; Fryer et al. 1999; in 't Zand 1998;
Ruffert \& Janka 1998, 1999; Woosley, MacFadyen, \& Heger 1999;  for a
discussion see also, e.g., M\'esz\'aros 1999a; Rees 1999).  In all
cases it is believed that the relic of the explosion is a
stellar-mass black hole. In other words, within this picture every
detected GRB (excluding SGR bursts) represents either a black hole
formation event in the universe, or at least a signal from a
pre-existing stellar-mass black hole. 

Although the central engine of a GRB is hidden from observation,
both classes of burst models make certain predictions that can be
tested by observations, mainly based on the ages of the objects
involved (cf. Bloom, Sigurdsson, \& Pols 1999; Fryer, Woosley, \& Hartmann 
1999). As noted by Paczy\'nski (1998b), massive stars will be close to
their birthplaces when they explode and hence they could be embedded
in a  dust-rich environment. Several pieces of evidence seem to 
favor this GRB model: \it First, \rm within the context of the
fireball models, multi-wavelength observations of afterglows can be
used to determine the gas density of the ``interstellar'' medium
surrounding the burster (cf. Vreeswijk et al. 1999; Wijers \& Galama
1999). Model fits led to the conclusion that the density of
the external medium into which the fireball of GRB 980329 expands
(see section \ref{afterglow}) is $\approx$1000 cm$^{-3}$ (Lamb et
al. 1999). In other words, this burst presumably occurred in an
interstellar gas cloud in a remote galaxy. (In contrary cases, e.g.
GRB 970508, an ambient gas density of about 1 cm$^{-3}$ or
less was deduced from the afterglow data  [Waxman 1997; Wijers \&
Galama 1999].) \it Second, \rm only about 50\% of all GRBs whose X-ray
afterglows were detected have been discovered in optical bands. This
could be due to extinction by dust in the GRB host galaxies. An
example of an optically obscured burst is GRB 970828. The burst
developed a bright X-ray afterglow detected
by \it RXTE \rm (Marshall et al. 1997; Remillard et al. 1997),
\it ASCA \rm (Murakami et al. 1997), and \it 
ROSAT \rm (Greiner et al. 1997b), but it was not seen in the optical
down to an $R$-band magnitude of 24 (Groot et al. 1998), although
follow-up
observations started only hours after the burst.\footnote{The only
near-infrared follow-up observations of this event were reported by
Klose, Eisl\"offel, \& Stecklum (1997), but they were performed 13
days after the burst. A red object was detected by these authors in
the 30$''$ \it ASCA \rm X-ray error circle (Murakami et al. 1997)
which was later found to be also located in the 10$''$ \it ROSAT \rm HRI error
circle (Greiner et al. 1997b). However, second epoch observations five
months later showed this object again (Klose 1998, unpublished). So,
it is not the GRB afterglow, despite it having an underlying faint galaxy
(Pedersen 1997, private communication).} \it Third, \rm  the GRB hosts
detected so far are actively star-forming galaxies (Fruchter et
al. 1999a). Their star-formation rates range between about 1
to 10 $M_\odot$ yr$^{-1}$ (cf. Bloom et al.  1998b,1999b;
Djorgovski et al. 1998; Kulkarni et
al. 1998a; Odewahn et al. 1998), although this is not unusual for galaxies
at these redshifts. 
There is presently no evidence for GRBs occurring in elliptical galaxies.

A number of future tests of the GRB environment have been proposed
in the literature.
For example, the gaseous component of the GRB environment could
manifest itself by certain spectral features  in the X-ray afterglow
(cf. B\"ottcher et al. 1999;  Lazzati, Campana, \& Ghisellini 1999;
M\'esz\'aros \& Rees 1999a; Piro et al. 1999b) or in the optical
afterglow (cf. Perna \& Loeb 1998a), whereas a test of the dusty
component could make use of the scattering properties of the dust
grains in the soft X-ray band (Klose 1998). 

\section{Evidence for a GRB - Supernova association \label{supernova}}

The idea of a GRB-Supernova (SN) association was first considered by Colgate
(1968, 1974; see also the review by Chupp 1976) but it was not
supported by observations at that time. About 30 years later it turned
out that a GRB-SN link does indeed exist.

GRB 980425 was detected by \it BeppoSAX \rm and an anomalous
SN (1998bw) was found in the 8 arcmin error circle  of the X-ray
afterglow (Galama et al. 1998; Kulkarni et al. 1998b). The SN spectrum
indicated  ejection speeds approaching 60\,000 km s$^{-1}$ (Kulkarni
et al. 1998b). SN 1998bw is classified as a peculiar Type Ic supernova. 
Its host galaxy is at a redshift of 0.0085 (Tinney et al. 1998),
corresponding to a distance of 38 Mpc ($H_0$ = 65 km s$^{-1}$
Mpc$^{-1}$). The gravitational collapse of a massive stellar iron
core that formed a black  hole can explain the observations (Iwamoto
et al. 1998; Woosley, Eastman, \& Schmidt 1999;  Woosley, MacFadyen,
\& Heger 1999). The  radio-to-X-ray light curves of this event is not
in conflict with the fireball model (Iwamoto 1999; for the optical
light curve see McKenzie \& Schaefer 1999). Since GRB 980425  was so
close to our Galaxy, but not exceptionally bright, one has to conclude
that \it BATSE \rm does only see the 'tip of the iceberg' of all GRBs
occurring in the universe. 

Although up to December 1999 no other direct GRB-SN association has been 
found, there is in fact increasing observational evidence that
at least a subclass of all bursts is associated with SNe. The
strongest evidence comes from the recent discovery that the light
curves of GRB 970228, 980326, and possibly 990712, show a bump at late
times which cannot be explained within the context of simple afterglow
models, but can be fitted as a redshifted SN 1998bw light curve which
adds to the flux of the GRB afterglow (Bloom et al. 1999a; Dar 1999;
Galama et al. 1999; Hjorth et al. 1999; Reichart 1999). If that
interpretation is correct, then GRB 970228, the first burst with a
detected optical afterglow, represents one of the most distant SNe
ever seen ($z$=0.695; Djorgovski et al. 1999). Moreover, if it turns
out that SN light curves do appear in GRB afterglows, this 
could represent a new powerful method to measure the
cosmological parameters. A check of this potential application might be an
interesting long-term project for 8-m class optical telescopes.

It has recently been proposed that SN 1997cy (Germany et al. 1999) and
SN 1999E (Kulkarni \& Frail 1999; Thorsett \& Hogg 1999), which seem
to resemble SN 1998bw, could be related to the \it BATSE \rm bursts
970514 and 980910a, respectively, which were
not seen by \it BeppoSAX \rm (for a discussion, see Wheeler 1999).  
It has also been suggested that there is 
observational evidence for a relativistic jet from SN 1987A (Cen
1999). Finally, the detection of a redshifted iron K$_\alpha$
emission line in the X-ray afterglows of GRB 970508 (Piro et
al. 1999a,b) and possibly GRB 970828 (Yoshida et al. 1999)
has been reported, which could
indicate an SN event before the  corresponding GRB (Lazzati,
Campana, \& Ghisellini 1999; Vietri \& Stella 1998). 
Naturally, because of these findings the question arises if all
GRBs are physically related to SNe.
Several authors have therefore compared the \it BATSE
\rm GRB catalog (Meegan et al. 1998) with SN catalogs (Bloom et
al. 1998a; Hudec, Hudcova, \& Hroch 1999;  Kippen et al. 1998b; Klose
1999; Norris, Bonnell, \& Watanabe 1999; Wang \& Wheeler 1998).
However, no convincing statistical evidence for an excess of any
subclass of SNe in GRB error boxes has been found in that way. The
reason for this is that \it BATSE\rm-only  error boxes are very large
in radius and that the SN database is very inhomogeneous. Therefore, these
studies cannot exclude the existence of a general association of GRBs
to SNe. In any case, GRB research gives a strong impulse to SN research
(Wheeler 1999; Woosley, MacFadyen, \& Heger 1999). If GRBs do indeed
reflect certain SN explosions, \it BATSE \rm detects more SNe per year than all
current SN search campaigns combined.

Based on these findings, is it possible that we have already seen a
few SN-GRBs in our Galactic neighborhood but not recognized them as
such? A search for potential GRB afterglows in the CBAT SN catalog has
not been successful (Klose 1999), although such a study is confronted
with the lack of detailed observational data about most known SNe.
Another way to tackle the question under consideration is to look  
for bright, nearby galaxies in arcmin-sized IPN error boxes. 
A working hypothesis can be  that a SN-GRB association was not
recognized in these cases because these GRB error boxes could not be determined
in a timely manner, i.e., any SN light curve was already below the
detection threshold when these error boxes were imaged in the
optical. A visual inspection of published optical images of
IPN error boxes of the 1970's 
(cf. Schaefer et al. 1998; Vrba, Hartmann, \& Jennings 1995) 
reveals one potential candidate for such a case, 
GRB 781104b.\footnote{Schaefer (1999) made a similar suggestion 
at the 5th Huntsville Symposium on GRBs, Huntsville, AL, October 1999.}
The only 14 arcmin$^2$ large error box of this burst contains a relatively
bright galaxy ($B$=15, Simbad data base). This galaxy, MCG 04-47-011,
is at a redshift of $z$ = 0.0024 (Simbad data base), corresponding to
a distance of about 11 Mpc ($H_0$ = 65 km s$^{-1}$ Mpc$^{-1}$). The
latter estimate is uncertain however, because the redshift is so
small. Similar to the duration of GRB 980425 (Galama et al. 1998a;
Pian et al. 1999), the duration of GRB 781104b was about 30 seconds
(Mazets et al. 1981; Teegarden \& Cline 1980). Although no SN is
known in MCG 04-47-011, another way to check whether
GRB 781104b was a nearby SN burst could be to search in the radio band for
its potential GRB remnant.

A number of authors has recently addressed the question of how the relics
of GRB explosions might look and where they are in our local
universe (e.g., Hansen 1999).  At a GRB rate in the order of 1 per
10$^{6\pm1}$ yr per Milky-Way like galaxy, GRB remnants in the local
universe  could possibly be detectable. Woods \& Loeb (1999) 
proposed to search for radio emission from GRB remnants in the Virgo
cluster of galaxies ($z$=0.0038; Simbad data base). This would require
surveying on the order of 1000 galaxies down to a flux level below 100
$\mu$Jy. Efremov, Elmegreen, \& Hodge (1998) as well as Loeb \& Perna
(1998b) brought attention to HI supershells (cf. Lee \& Irwin
1997; Walter et al. 1998), whose energy content is comparable to the 
electromagnetic energy release of GRBs. Potential X-ray selected
GRB remnant candidates in M\,101 have also been proposed (Wang
1999). First calculations of the emission spectrum of GRB remnants
have been published and phenomenological differences to supernova
remnants (SNRs) have been outlined (Perna, Raymond, \& Loeb
1999). Since there are some similarities to SNRs, however, in the
optical the search strategy for GRB remnant candidates could follow
those for HII-regions/SNRs (cf.  Elmegreen \& Salzer 1999; Gordon et
al. 1993; Matonick \& Fesen 1997). The close-by, face-on spirals
M\,33 and NGC\,300 could be potential targets for such an
investigation.

\section{The future of GRB research}

\it BeppoSAX \rm currently detects about 10 X-ray afterglows  of GRBs
per year. The \it HETE-2 \rm satellite (Vanderspek et al. 1999),
scheduled to be launched early 2000, will increase this rate by a
factor of 3 to 5, and the \it Swift \rm GRB mission (Gehrels 1999),
scheduled to fly in 2003...2005, will provide about 100 or more very
small GRB error boxes per year. This, combined with the next generation
high-energy satellites (\it XMM, AXAF-Chandra, ASTRO-E, Integral\rm),
will guarantee a large increase in our knowledge about GRBs in the
coming decade (see also Hurley 1999b).

For optical telescopes there are three observing strategies: 
1) \it Imaging of GRB error boxes when the
burster is still active in the gamma-ray band. \rm GRB
990123 has already demonstrated that this is feasible with current CCD
cameras. Great progress will be achieved when automatic 1-m class
telescopes will become available, like the Flagstaff 1.3-m (Vrba
et al. 1999b) and the Super-Lotis 0.6-m (Park et al. 1999). Prism
spectroscopy is another future option. 2) \it Investigation of GRB
afterglows. \rm In most cases this will require 'Target of
Opportunity' observations scheduled within hours or days after a burst. 
3) \it Study of GRB host galaxies. \rm This can be performed even months
after a burst. It requires the largest telescopes.

Some open questions for the coming decade are obvious: Are all GRBs
related to SNe? What is the physical difference between the short and
the long bursts? How many distinct populations of bursters are there? 
Do GRBs occur in elliptical galaxies? Do short bursts develop afterglows?
Does any subclass of GRBs represent a cosmological
standard candle?  Can we measure GRB redshifts based on observations
in the gamma-ray band alone? What are the most distant GRBs? Where are 
the GRB remnants in the local universe, in our Galaxy, and in M\,31? 
Finally, do GRBs affect the evolution of life? 

GRBs are the most energetic electromagnetic phenomena in the
universe. Although for an individual GRB this holds only for a very
small timespan, it can make GRBs a giant observational tool to
investigate the high-$z$ universe (cf. Lamb \& Reichart 1999). This
includes the cosmological parameters (cf. Holz, Miller, \& Quashnock
1999; Horack et al. 1996; Marani et al. 1999), the cosmic radiation
background (Mannheim, Hartmann, \& Funk 1996), stellar evolution, the
cosmic star formation rate (Hartmann \& Band 1998; Jorgensen et al. 1995; 
Krumholz, Thorsett, \& Harrison 1998;
Mao \& Mo 1998; Totani 1999; Wijers et al. 1998;), large-scale
structure (Hartmann \& Blumenthal 1989; Lamb \& Quashnock 1993), etc. 
GRBs could also prove to be
a useful tool to investigate the interstellar medium in our Galaxy
(and in intergalactic space) through which they are observed.
Thus, there are good reasons to believe that the future
of GRB research is bright, in all bands of the electromagnetic spectrum.

\vspace*{10mm}

\noindent \bf Appendix: selected WWW-pages about GRBs \rm

\vspace*{2mm}

\begin{itemize}
\itemsep-1mm
\item GRB coordinated network GCN: http://gcn.gsfc.nasa.gov/gcn/
\item GRB missions: Holger Pedersen's WWW-page at \\
      http://www.astro.ku.dk/$^\sim$holger/ddirGAMMA/dF/OFSAT.html
\item SGRs: Robert Duncan's WWW-page at the University of Texas Austin, \\ 
      http://solomon.as.utexas.edu/$^\sim$duncan/
\item IPN: Kevin Hurley's WWW-page at Berkeley, \\
      http://ssl.berkeley.edu/ipn3/interpla.html
\item GRB bibliography: Kevin Hurley's WWW-page at Berkeley, \\
      http://ssl.berkeley.edu/ipn3/bibliogr.html
\item reports about GRB follow-up observations: 
      Jochen Greiner's WWW-page at the AIP,
      http://www.aip.de/$^\sim$jcg/grbgen.html
\item \it BATSE\rm: http://gammaray.msfc.nasa.gov/batse/
\item \it BeppoSAX\rm: http://www.sdc.asi.it/
\item \it HETE-2 \rm satellite: http://space.mit.edu/HETE/
\item \it Swift \rm mission: http://swift.gsfc.nasa.gov/
\end{itemize}

\vspace*{2mm}


\vspace*{5mm}

\noindent \bf Acknowledgements \rm

\vspace*{2mm}

I am deeply indebted to Bringfried Stecklum (Th\"uringer
Landessternwarte Tautenburg) and Olaf Fischer
(Universit\"ats-Sternwarte Jena), who share my interest in
GRBs and their afterglows. I am grateful to Jochen Greiner (Astrophysical
Institute Potsdam), Dieter Hartmann (Clemson University,  Clemson,
SC), Attila M\'esz\'aros (Astronomical Institute, Charles University
Prague), Bringfried Stecklum, and Fred Vrba (U.S. Naval Observatory,
Flagstaff, AZ)  for many valuable comments on the manuscript. I thank
Josef Solf for his interest in and support of GRB research during his
tenure as director of the Th\"uringer Landessternwarte Tautenburg
in the years 1994-99. I greatly appreciate assistance during observing
campaigns at the Calar Alto Observatory, Spain, from Kurt Birkle,
Ulrich Thiele, and Markus Feldt, Max-Planck-Institut f\"ur Astronomie,
Heidelberg. This research has made use of the Simbad database,
operated at CDS, Strasbourg, France.


\vspace{0.7cm}
\noindent
\large{\bf References}

\small
\begin{description}
\itemsep-0.5mm
\itemindent-1.03cm

\item Akerlof, C. et al., 1995, Astroph. Sp. Sci. 231, 255
\item Akerlof, C. et al., 1999, Nature 398, 400
\item Andersen, M. I. et al., 1999, Science 283, 2075
\item Bal\'azs, L. G., M\'esz\'aros, A., \& Horv\'ath, I., 1998, A\&A 339, 1
\item Bal\'azs, L. G. et al., 1999, A\&A Suppl. Ser. 138, 417
\item Baring, M. G. \& Harding, A. K., 1998, Adv. Space Res. 22 (7), 1115
\item Barthelmy, S. D. et al., 1998,
           in: C. A. Meegan, R. D. Preece, \& T. M. Koshut (eds.),
           4th Huntsville Symposium on Gamma-Ray Bursts
           (AIP: New York), AIP Conf. Proc. 428, 99
\item Beuermann, K. et al., 1999, A\&A 352, L26
\item Blaes, O. M., 1994, ApJ Suppl. Ser. 92, 643
\item Blaes, O. et al., 1997, ApJ 479, 868
\item Blinnikov, S., 1999, preprint astro-ph/9911138
\item Bloom, J. S., Sigurdsson, S., \& Pols, O. R., 1999, MNRAS 305, 763
\item Bloom, J. S. et al., 1998a, ApJ 506, L105 
\item Bloom, J. S. et al., 1998b, ApJ 507, L25 
\item Bloom, J. S. et al., 1999a, Nature 401, 453
\item Bloom, J. S. et al., 1999b, ApJ 518, L1
\item Boella, G. et al., 1997, A\&A Suppl. Ser. 122, 299
\item Bo\"er, M. et al., 1993, A\&A 277, 503
\item Bo\"er, M. et al., 1997, ApJ 481, L39
\item Bond, H. E., 1997, IAU Circ. 6654
\item Bonnell, J. T. \& Klebesadel, R. W., 1996,
           in: C. Kouveliotou, M. F. Briggs, \& G. J. Fishman (eds.),
           3rd Huntsville Symposium on Gamma-Ray Bursts
           (AIP: New York), AIP Conf. Proc. 384, 977
\item B\"ottcher, M. et al., 1999, A\&A 343, 111
\item Bradt, H. V. \& Smith, D. A., 1999, A\&A Suppl. Ser. 138, 423
\item Brainerd, J. J., 1994, ApJ 428, L1
\item Briggs, M. et al., 1996, ApJ 459, 40
\item Briggs, M. et al., 1999, ApJ Suppl. Ser. 122, 503
\item Burenin, R. A. et al., 1998, Astron. Lett. 24, 427; astro-ph/9804274
\item Castro-Tirado, A. J., 1999, Astroph. Sp. Sci. 263, 15; astro-ph/9903187
\item Castro-Tirado, A. J. et al., 1997, Science 279, 1011
\item Castro-Tirado, A. J. et al., 1999, Science 283, 2069
\item Cavallo, G. \& Rees, M. J., 1978, MNRAS 183, 359
\item Cen, R., 1999, ApJ 524, L51
\item Chevalier, R. A. \& Li, Z.-Y., 1999, preprint astro-ph/9908272
\item Chupp, E. L., 1976, Gamma Ray Astronomy
           (Reidel: Dordrecht, Boston)
\item Cline, D. B., Matthey, C., \& Otwinowski, S., 1999, preprint
           astro-ph/9905346
\item Cline, T. L. et al., 1980, ApJ 237, L1
\item Colgate, S. A., 1968, Canadian J. Phys. 46, S476
\item Colgate, S. A., 1974, ApJ 187, 333
\item Costa, E., 1999, talk given at the 5th
           Huntsville Symposium on  Gamma-Ray Bursts, AIP Conf. Proc., 
           to be published
\item Costa, E. et al., 1997a, IAU Circ. 6533
\item Costa, E. et al., 1997b, IAU Circ. 6572
\item Costa, E. et al., 1997c, IAU Circ. 6576
\item Costa, E. et al., 1998, Adv. Space Res. 22 (7), 1129
\item Covino, S. et al., 1999, A\&A 348, L1
\item Dar, A., 1999, GCN 346
\item Dermer, C. D. \& Weiler, T. J., 1995, Astroph. Sp. Sci. 231, 377
\item Dezalay, J. P. et al., 1996, ApJ 471, L27
\item Djorgovski, S. G. et al., 1998, ApJ 508, L17
\item Djorgovski, S. G. et al., 1999, GCN 289
\item Draine, B. T., 1999, ApJ, submitted, preprint astro-ph/9907232
\item Efremov, Y. N., Elmegreen, B. G., \& Hodge, P. W., 1998, ApJ 501, L163
\item Elmegreen, D. M. \& Salzer, J. J., 1999, ApJ Suppl. Ser. 117, 764
\item Fenimore, E. E. \& Bloom, J. S., 1995, ApJ 453, 25
\item Fenimore, E. E. et al., 1993, Nature 366, 40
\item Fishman, G. J., 1981, Astroph. Sp. Sci. 75, 125
\item Fishman, G. J., 1995, PASP 107, 1145
\item Fishman, G. J., 1999, A\&A Suppl. Ser. 138, 395
\item Fishman, G. J. \& Inan, U. S., 1988, Nature 331, 418
\item Fishman, G. J. \& Meegan, C. A., 1995, 
      Annu. Rev. Astron. Astrophys. 33, 415
\item Fishman, G. J. et al., 1993, A\&A Suppl. Ser. 97, 17
\item Fishman, G. J. et al., 1994, ApJ Suppl. Ser. 92, 229
\item Ford, L. A. \& Band, D. L., 1996, ApJ 473, 1013
\item Frail, D. A., 1999, talk given at the 5th
           Huntsville Symposium on  Gamma-Ray Bursts, AIP Conf. Proc.,
           to be published
\item Frail, D. A. et al., 1994, ApJ 437, L43
\item Frail, D. A. et al., 1997a, ApJ 483, L91
\item Frail, D. A. et al., 1997b, Nature 389, 261
\item Frail, D. A. et al., 1999a, GCN 451
\item Frail, D. A. et al., 1999b, ApJ 525, L81
\item Frontera, F., 1998, in:
           M. S. Potgieter, B. C. Raubenheimer, \& D. J. van der Wait (eds.),
           25th International Cosmic Ray Conference, Vol. 8,
           (World Scientific: Singapore), p. 307; astro-ph/9802157
\item Fruchter, A., 1999, ApJ 512, L1
\item Fruchter, A. et al., 1999a, ApJ 516, 683
\item Fruchter, A. et al., 1999b, preprint astro-ph/9903236
\item Fryer, C. L. \& Woosley, S. E., 1998, ApJ 502, L9
\item Fryer, C. L., Woosley, S. E., \& Hartmann, D. H., 1999, ApJ 526, 152
\item Fryer, C. L. et al., 1999, ApJ 520, 650
\item Galama, T. J. et al., 1997a, A\&A 321, 229
\item Galama, T. J. et al., 1997b, ApJ 486, L5
\item Galama, T. J. et al., 1998, Nature 395, 670
\item Galama, T. J. et al., 1999, preprint astro-ph/9907264 
\item Gehrels, N. 1999, talk given at the 5th
           Huntsville Symposium on Gamma-Ray Bursts,
           AIP Conf. Proc., to be published
\item Germany, L. M. et al., 1999, preprint astro-ph/9906096
\item Ghisellini, G., 1999, preprint astro-ph/9907376
\item Ghisellini, G. \& Lazzati, D., 1999, MNRAS 309, L7
\item Giblin, T. W. et al., 1999, ApJ 524, L47
\item Goodman, J., 1997, New Astron. 2, 449
\item Gordon, S. M. et al., 1993, ApJ 418, 743
\item Gorosabel, J. et al., 1995, Astroph. Sp. Sci. 231, 297
\item Greiner, J., 1995, Astroph. Sp. Sci. 231, 263
\item Greiner, J., 1998, in:
      Proc. Multifrequency behaviour of high energy 
      cosmic sources, Vulcano, Mem. Societa Astron. Italiana,
      in press; preprint astro-ph/9802222
\item Greiner, J. \& Motch, C., 1995, A\&A 294, 177
\item Greiner, J. et al., 1991, in:
      Proc. 22nd International Cosmic Ray Conference
      (Reprint Ltd: Dublin), ISBN 1 85500 995 1, Vol. 1, 53
\item Greiner, J. et al., 1992, in:
      W. Paciesas \& G. Fishman (eds.), Gamma-Ray Bursts (AIP: New York), 
      AIP Conf. Proc. 265, 327
\item Greiner, J. et al., 1996, in:
           C. Kouveliotou, M. F. Briggs, \& G. J. Fishman (eds.),
           3rd Huntsville Symposium on Gamma-Ray Bursts
           (AIP: New York), AIP Conf. Proc. 384, 622
\item Greiner, J. et al., 1997a, A\&A 325, 640
\item Greiner, J. et al., 1997b, IAU Circ. 6757
\item Greiner, J. et al., 1999a, A\&A Suppl. Ser. 138, 441
\item Greiner, J. et al., 1999b, preprint astro-ph/9910300
\item Grindlay, J. E. 1999, ApJ 510, 710
\item Groot, P. J. et al., 1997, IAU Circ. 6584
\item Groot, P. J. et al., 1998, ApJ 493, L27
\item Gruzinov, A., 1999, ApJ 525, L29
\item Gruzinov, A. \& Waxman, E., 1999, ApJ 511, 852
\item Hakkila, J. et al., 1994, ApJ 422, 659
\item Halpern, J. P. et al., 1999, ApJ 517, L105
\item Hansen, B. M., 1999, ApJ 512, L117
\item Harding, A. K., 1994, ApJ Suppl. Ser. 90, 863
\item Harrison, F. A. et al. 1999, ApJ 523, L121
\item Hartmann, D. H., 1994, Science 263, 47 
\item Hartmann, D. H., 1995, A\&A Rev. 6, 225
\item Hartmann, D. H., 1996, A\&A Suppl. Ser. 120, 31
\item Hartmann, D. H., 1999, Perspectives of the National Acad. of 
      Sciences (PNAS) 96, 4752
\item Hartmann, D. H. \& Band, D. L., 1998, in:
           C. A. Meegan, R. D. Preece, \& T. M. Koshut (eds.),
           4th Huntsville Symposium on Gamma-Ray Bursts
           (AIP: New York), AIP Conf. Proc. 428, 615 
\item Hartmann, D. H. \& Blumenthal, G. R., 1989, ApJ 342, 521
\item Hartmann, D. H. \& Narayan, R., 1996, ApJ 464, 226
\item Higdon, J. C. \& Lingenfelter, R. E., 1990, Annu. Rev. Astron.
           Astrophys. 28, 401
\item Hjorth, J. et al., 1999, GCN 403
\item Holz, D. E., Miller, M. C., \& Quashnock, J. M., 1999, ApJ 510, 54
\item Horack, J. M., Mallozzi, R. S., \& Koshut, T. M., 1996, ApJ 466, 21
\item Horack, J. M. et al., 1996, ApJ 472, 25
\item Horv\'ath, I., 1998, ApJ 508, 757
\item Horv\'ath, I., M\'esz\'aros, P., \& M\'esz\'aros, A., 1996, ApJ 470, 56
\item Houston, B. P. \& Wolfendale, A. W., 1983, Vistas Astron. 26, 107
\item Hudec, R., 1993, A\&A Suppl. Ser. 97, 49
\item Hudec, R. \& Wenzel, W., 1996, A\&A Suppl. Ser. 120, 707
\item Hudec, R., Hudcova, V., \& Hroch, F. 1999, 138, 475
\item Hudec, R. et al., 1996, in:
      H. R. Miller, J. R. Webb, \& J. C. Noble (eds.),
      Blazar Continuum Variability (ASP: San Francisco),
      ASP Conf. Ser. 110, 129
\item Hughes, P. A., Aller, H. D., \& Aller, M. F., 1985, ApJ 298, 301
\item Hurley, K., 1994, ApJ Suppl. Ser. 90, 857
\item Hurley, K., 1998a, in:
           C. A. Meegan, R. D. Preece, \& T. M. Koshut (eds.),
           4th Huntsville Symposium on Gamma-Ray Bursts
           (AIP: New York), AIP Conf. Proc. 428, 87
\item Hurley, K., 1998b, in: Proc.
      3rd Integral Workshop, preprint astro-ph/9812052
\item Hurley, K., 1999a, preprint astro-ph/9912061
\item Hurley, K., 1999b, A\&A Suppl. Ser. 138, 553
\item Hurley, K. et al., 1994, ApJ Suppl. Ser. 92, 655
\item Hurley, K. et al., 1995, Astroph. Sp. Sci. 231, 227
\item Hurley, K. et al., 1999a, ApJ 515, 497
\item Hurley, K. et al., 1999b, ApJ 524, 92
\item in 't Zand, J. J. M., 1998, ApJ 503, L53 
\item in 't Zand, J. J. M. et al., 1998, ApJ 505, L119
\item Israel, G. L. et al., 1999, A\&A 348, L5
\item Iwamoto, K., 1999, ApJ 512, L47
\item Iwamoto, K. et al., 1998, Nature 395, 672
\item Jorgensen, H. et al., 1995, Astroph. Sp. Sci. 231, 389
\item Katz, J. I. \& Canel, L. M., 1996, ApJ 471, 915
\item Kippen, R. M. et al., 1998a,
           in: C. A. Meegan, R. D. Preece, \& T. M. Koshut (eds.),
           4th Huntsville Symposium on Gamma-Ray Bursts
           (AIP: New York), AIP Conf. Proc. 428, 119
\item Kippen, R. M. et al., 1998b, ApJ 506, L27
\item Klebesadel, R. W., Strong, I. B., \& Olson, R. A., 1973, ApJ 182, L85
\item Klose, S., 1994, ApJ 423, L23
\item Klose, S., 1995a, A\&A 303, 75
\item Klose, S., 1995b, ApJ 446, 261
\item Klose, S., 1995c, ApJ 446, 357
\item Klose, S., 1998, ApJ 507, 300
\item Klose, S., 1999, in: Proc. 19th Texas Symposium on
           Relativistic Astrophysics, Paris, France, submitted
\item Klose, S., Eisl\"offel, J., \& Richter, S., 1996, ApJ 470, L93
\item Klose, S., Eisl\"offel, J., \& Stecklum, B., 1997, IAU Circ. 6756
\item Klose, S., Meusinger, H., \& Lehmann, H., 1998, IAU Circ. 6864
\item Klose, S., Stecklum, B., \& Fischer, O., 1999,
           Proc. 5th Huntsville Symposium on Gamma-Ray Bursts,
           AIP Conf. Proc., submitted
\item Klose, S. et al., 1998,
           in: C. A. Meegan, R. D. Preece, \& T. M. Koshut (eds.),
           4th Huntsville Symposium on Gamma-Ray Bursts
           (AIP: New York), AIP Conf. Proc. 428, 635
\item Kouveliotou, C., 1995a, in:
           H. B\"ohringer, G. E. Morfill, \& J. E. Tr\"umper (eds.),
           Proc. 17th Texas Symposium on Relativistic
           Astrophysics (The New York Academy of Sciences: New York), 
           Ann. N.Y. Acad. Sci. 759, 411
\item Kouveliotou, C., 1995b, Astroph. Sp. Sci. 231, 49
\item Kouveliotou, C. et al., 1993, ApJ 413, L101
\item Krumholz, M., Thorsett, S. E., \& Harrison, F. A., 1998, ApJ 506, L81
\item Kulkarni, S. R. \& Frail, D. A., 1999, GCN 198
\item Kulkarni, S. R. et al., 1998a, Nature 393, 35 
\item Kulkarni, S. R. et al., 1998b, Nature 395, 663
\item Kulkarni, S. R. et al., 1999, Nature 398, 389
\item Lamb, D. Q., 1995, PASP 107, 1152
\item Lamb, D. Q., 1997, in: Gravitation, R. E. Schielicke (ed.),
            Rev. Mod. Astron. 10, 101       
\item Lamb, D. Q., 1999, A\&A Suppl. Ser. 138, 607
\item Lamb, D. Q. \& Quashnock, J. M., 1993, 415, L1
\item Lamb, D. Q. \& Reichart, D., 1999, preprint astro-ph/9909002
\item Lamb, D. Q. et al., 1999, A\&A Suppl. Ser. 138, 479
\item Laros, J. G. et al., 1986, Nature 322, 152
\item Laros, J. G. et al., 1987, ApJ 320, L111
\item Laros, J. G. et al., 1997, ApJ Suppl. Ser. 110, 157
\item Laros, J. G. et al., 1998, ApJ Suppl. Ser. 118, 391
\item Larson, S. B. \& McLean, I. S., 1997, ApJ 491, 93
\item Larson, S. B., McLean, I. S., \& Becklin, E. E., 1996, ApJ, 460, L95
\item Lazzati, D., Campana, S., \& Ghisellini, G., 1999, MNRAS 304, L31
\item Lee, B. et al., 1997, ApJ 482, L125
\item Lee, S.-W. \& Irwin, J. A., 1997, ApJ 490, 247
\item Lee, T. T. \& Petrosian, V., 1997, ApJ 474, 37
\item Li, H. \& Liang, E. P., 1992, ApJ 400, L59
\item Loeb, A. \& Perna, R., 1998a, ApJ 495, 597
\item Loeb, A. \& Perna, R., 1998b, ApJ 503, L35
\item Luginbuhl, C. et al., 1995, Astroph. Sp. Sci. 231, 289
\item Luginbuhl, C. et al., 1996, IAU Circ. 6526
\item Lund, N., 1995, Astroph. Sp. Sci. 231, 217
\item Mannheim, K., Hartmann, D. H., \& Funk, B., 1996, ApJ 467, 532
\item Mao, S. \& Mo, H.-J., 1998, A\&A 339, L1
\item Mao, S., Narayan, R., \& Piran, T., 1994, ApJ 420, 171
\item Mao, S. \& Paczy\'nski, B, 1992a, ApJ 388, L45
\item Mao, S. \& Paczy\'nski, B, 1992b, ApJ 389, L13
\item Marani, G. F. et al., 1999, ApJ 512, L13
\item Marshall, F. E. et al., 1997, IAU Circ. 6727
\item Matonick, D. M. \& Fesen, R. A., 1997, ApJ Suppl. Ser. 112, 49
\item Mazets, E. P. et al., 1981, Astroph. Sp. Sci. 80, 3
\item McBreen, B. et al., 1994, MNRAS 271, 662
\item McKenzie, E. H. \& Schaefer, B. E., 1999, PASP 111, 964
\item McNamara, B. J. \& Harrison, T. E., 1998, Nature 396, 233
\item McNamara, B. J., Harrison, T. E., \& Williams, C. L., 1995, ApJ 452, L25
\item Meegan, C. A., 1998, Adv. Space Res. 22 (7), 1065
\item Meegan, C. A. et al., 1992, Nature 355, 143
\item Meegan, C. A. et al., 1995, ApJ 446, L15
\item M\'esz\'aros, A., Bagoly, Z., \& Vavrek, R., 1999, preprint 
      astro-ph/9912037
\item M\'esz\'aros, A. \& M\'esz\'aros, P., 1996, ApJ 466, 29
\item M\'esz\'aros, A. et al., 1996, J. Korean Astron. Soc. 29, S43
\item M\'esz\'aros, P., 1997, in: Gravitation, R. E. Schielicke (ed.),
      Rev. Mod. Astron. 10, 127 
\item M\'esz\'aros, P., 1999a, preprint astro-ph/9904038
\item M\'esz\'aros, P., 1999b, Nature 398, 368
\item M\'esz\'aros, P., 1999c, A\&A Suppl. Ser. 138, 533
\item M\'esz\'aros, P. \& M\'esz\'aros, A., 1995, ApJ 449, 9 
\item M\'esz\'aros, P. \& Rees, M. J., 1999a, ApJ 502, L105
\item M\'esz\'aros, P. \& Rees, M. J., 1999b, MNRAS 306, L39
\item M\'esz\'aros, P., Rees, M. J., \& Papathanassiou, H., 1994, ApJ 432, 181
\item M\'esz\'aros, P., Rees, M. J., \& Wijers, R. A. M. J., 1998, ApJ 499, 301
\item M\'esz\'aros, P., Rees, M. J., \& Wijers, R. A. M. J.,
           1999, New Astron. 4, 303
\item Mitrofanov, I. G., 1998, Adv. Space Res. 22 (7), 1077
\item Mukherjee, S. et al., 1998, ApJ 508, 314
\item Murakami, T. et al., 1997, IAU Circ. 6732
\item Nemiroff, R. J., 1994, Comm. Astrophys. 17, 189
\item Nemiroff, R. J., 1995, PASP 107, 1131
\item Norris, J. P., Bonnell, J. T., \& Watanabe, K., 1999, ApJ 518, 901 
\item Norris, J. P. et al., 1994, ApJ 424, 540
\item Norris, J. P. et al., 1995, ApJ 439, 542
\item Odewahn, S. C. et al., 1998, ApJ 509, L5
\item Owens, A., Schaefer, B. E., \& Sembay, S., 1995, ApJ 447, 279
\item Paczy\'nski, B., 1987, ApJ 317, L51
\item Paczy\'nski, B., 1991a, Acta Astron. 41, 157
\item Paczy\'nski, B., 1991b, Acta Astron. 41, 257
\item Paczy\'nski, B., 1995, PASP 107, 1167
\item Paczy\'nski, B., 1998a,
           in: C. A. Meegan, R. D. Preece, \& T. M. Koshut (eds.),
           4th Huntsville Symposium on Gamma-Ray Bursts
           (AIP: New York), AIP Conf. Proc. 428, 783
\item Paczy\'nski, B., 1998b, ApJ 494, L45
\item Paczy\'nski, B., 1999, in:
      M. Livio, K. Sahu, \& N. Panagia (eds.), The Largest
      Explosions Since the Big Bang: Supernovae and Gamma-Ray Bursts
      (Cambridge University Press: Cambridge), preprint astro-ph/9909048
\item Palazzi, E. et al., 1998, A\&A 336, L95
\item Palmer, D. M. et al., 1995, Astroph. Sp. Sci. 231, 315
\item Panaitescu, A., M\'esz\'aros, P., \& Rees, M. J., 1998, ApJ 503, 314
\item Park, H. S. et al., 1997, ApJ 490, 99
\item Park, H. S. et al., 1999,
           Proc. 5th Huntsville Symposium on Gamma-Ray Bursts,
           AIP Conf. Proc., to appear
\item Pedersen, H., 1994, A\&A 291, L17
\item Pedersen, H. et al., 1998, ApJ 496, 311
\item Perna, R. \& Loeb, A., 1998a, ApJ 501, 467
\item Perna, R. \& Loeb, A., 1998b, ApJ 509, L85
\item Perna, R., Raymond, J., \& Loeb, A., 1999, preprint astro-ph/9904181
\item Pian, E. et al., 1999, preprint astro-ph/9903113
\item Piran, T., 1997, in: J. N. Bahcall \& J. P. Ostriker (eds.),
      Unsolved Problems in Astrophysics (Princeton University Press:
      Princeton), p. 343
\item Piran, T., 1999a, Physics Reports 314, 575; astro-ph/9810256
\item Piran, T., 1999b, preprint astro-ph/9907392
\item Piro, L. et al., 1999a, preprint astro-ph/9906363
\item Piro, L. et al., 1999b, ApJ 514, L73
\item Pizzichini, G. et al., 1986, ApJ 301, 641
\item Puget, J.-L., 1981, Astroph. Sp. Sci. 75, 109
\item Pugliese, G., Falcke, H., \& Biermann, P. L., 1999, A\&A 344, L37
\item Rees, M. J., 1995, PASP 107, 1176
\item Rees, M. J., 1998, in:
           A. V. Olinto, J. A. Frieman, \& D. N. Schramm (eds.),
           Proc. 18th Texas Symposium on Relativistic
           Astrophysics (World Scientific: Singapore), p. 34
\item Rees, M. J., 1999, A\&A Suppl. Ser. 138, 491
\item Reichart, D., 1999, ApJ 521, L111
\item Reichart, D. et al., 1999, ApJ 517, 692
\item Remillard, R. et al., 1997, IAU Circ. 6726
\item Rhoads, J. E., 1997, ApJ 487, L1
\item Rhoads, J. E., 1999, A\&A Suppl. Ser. 138, 539
\item Rol, E. et al., 1999,
           Proc. 5th Huntsville Symposium on Gamma-Ray Bursts,
           AIP Conf. Proc., to appear
\item Ruderman, M., 1975, Ann. N.Y. Acad. Sci. 262, 164
\item Ruffert, M. \& Janka, H.-Th., 1998, A\&A 338, 535
\item Ruffert, M. \& Janka, H.-Th., 1999, A\&A 344, 573
\item Sari, R., 1999, ApJ 524, L43
\item Sari, R. \& Piran, T., 1999, ApJ 517, L109
\item Sari, R., Piran, T., \& Halpern, J. P., 1999, ApJ 519, L17
\item Sari, R., Piran, T., \& Narayan, R., 1998, ApJ 497, L17
\item Schaefer, B. E., 1990, ApJ 364, 590
\item Schaefer, B. E., 1999,
           Proc. 5th Huntsville Symposium on Gamma-Ray Bursts,
           AIP Conf. Proc., to appear
\item Schaefer, B. E. et al., 1989, ApJ 340, 455
\item Schaefer, B. E. et al., 1994, ApJ 422, L71
\item Schaefer, B. E. et al., 1997, ApJ 489, 693
\item Schaefer, B. E. et al., 1998, ApJ Suppl. Ser. 118, 353
\item Schaefer, B. E. et al., 1999, ApJ 524, L103 
\item Schartel, N., Andernach, H., \& Greiner, J., 1997, A\&A 323, 659
\item Schmidt, M., 1999, ApJ 523, L117
\item Smith, D. A. et al., 1999, ApJ 526, 683
\item Smith, I. A. et al., 1999, A\&A 347, 92
\item Stanek, K. Z. et al., 1999, ApJ 522, L39
\item Stern, B., 1999, in: 
      J. Poutanen \& R. Svensson (eds),
      High Energy Processes in Accreting Black Holes
      (ASP: San Francisco), ASP Conf. Ser. 161, 277   
\item Strong, I. B., Klebesadel, R. W., \& Olson, R. A., 1974, ApJ 188, L1
\item Tavani, M., 1998, ApJ 497, L21
\item Taylor, G. B. et al., 1998a, GCN 40
\item Taylor, G. B. et al., 1998b, ApJ 502, L115
\item Teegarden, B. J., 1998, Adv. Space Res. 22 (7), 1083
\item Teegarden, B. J. \& Cline, T. L., 1980, ApJ 236, L67
\item Tegmark, M. et al., 1996a, ApJ 466, 757
\item Tegmark, M. et al., 1996b, ApJ 468, 214 
\item Thorsett, S. E. \& Hogg, D. W., 1999, GCN 197
\item Tinney, C. et al., 1998, IAU Circ. 6896
\item Totani, T. 1999, ApJ 511, 41
\item Trimble, V., 1995, PASP 107, 1133
\item Usov, V. V. \& Chibisov, G. V., 1975, Sov. Astron. 19, 115
\item van den Bergh, S., 1983, Astroph. Sp. Sci. 97, 385
\item van Paradijs, J. et al., 1997, Nature 386, 686
\item Vanderspek, R., Krimm, H. A., \& Ricker, G. R., 1995, Astroph. Sp. 
      Sci. 231, 259
\item Vanderspek, R. et al., 1999, A\&A Suppl. Ser. 138, 565
\item Vietri, M., 1999, preprint astro-ph/9911523
\item Vietri, M. \& Stella, L., 1998, ApJ 507, L45
\item Vrba, F. J., 1996,
           in: C. Kouveliotou, M. F. Briggs, \& G. J. Fishman (eds.),
           3rd Huntsville Symposium on Gamma-Ray Bursts
           (AIP: New York), AIP Conf. Proc. 384, 565
\item Vrba, F. J., Hartmann, D. H., \& Jennings, M. C., 1995, ApJ 446, 115
\item Vrba, F. J. et al., 1994, ApJ 424, 68
\item Vrba, F. J. et al., 1999a, ApJ 511, 298
\item Vrba, F. J. et al., 1999b,
           Proc. 5th Huntsville Symposium on Gamma-Ray Bursts,
           AIP Conf. Proc., to appear
\item Vreeswijk, P. M. et al., 1999, ApJ 523, 171
\item Wang, D. Q., 1999, ApJ 517, L27
\item Wang, L. \& Wheeler, J. C., 1998, ApJ 504, L87
\item Walter, F. et al., 1998, ApJ 502, L143
\item Waxman, E., 1997, ApJ 489, L33
\item Waxman, E., Kulkarni, S. R., \& Frail, D. A., 1998, ApJ 497, 288
\item Webber, W. R. et al., 1995, AJ 110, 733
\item Wei, D. M. \& Lu, T., 1999, preprint astro-ph/9908273
\item Wheeler, C., 1999, in:
      M. Livio, K. Sahu, \& N. Panagia (eds.), The Largest
      Explosions Since the Big Bang: Supernovae and Gamma-Ray Bursts
      (Cambridge University Press: Cambridge), preprint astro-ph/9909096
\item Wijers, R. A. M. J. \& Galama, T. J., 1999, ApJ 523, 177
\item Wijers, R. A. M. J., Rees, M. J., \& M\'esz\'aros, P., 1997,
      MNRAS 288, L51
\item Wijers, R. A. M. J. et al., 1998, MNRAS 294, L13
\item Wijers, R. A. M. J. et al., 1999, ApJ 523, L33
\item Woods, E. \& Loeb, A., 1999, preprint astro-ph/9907110
\item Woosley, S. E., Eastman, R. G., \& Schmidt, B. P., 1999, ApJ 516, 788
\item Woosley, S. E., MacFadyen, A. I., \& Heger, A., 1999, in:
      M. Livio, K. Sahu, \& N. Panagia (eds.), The Largest
      Explosions Since the Big Bang: Supernovae and Gamma-Ray Bursts
      (Cambridge University Press: Cambridge), preprint astro-ph/9909034
\item Yoshida, A. et al., 1999, A\&A Suppl. Ser. 138, 433
\item Zharikov, S. V. \& Sokolov, V. V., 1999, A\&A Suppl. Ser. 138, 485
\end{description}

\vfill

\end{document}